\documentclass[12pt,epsfig]{iopart}

\usepackage{graphics}
\usepackage{epsfig}
\begin{document}

\title[SQM2006 Xiaorong Wang ]{Measurement of Open Heavy Flavor Production with Single Muons in p+p and d+Au Collisions at RHIC}

\author{Xiaorong Wang, for PHENIX collaboration}

\address{New Mexico State University, Las Cruces, NM 88003 USA\\
Hua-zhong Normal University, Wuhan 430079} \ead{xrwang@bnl.gov}

\begin{abstract}
Heavy flavor production in hadronic collisions is dominated by
gluonic processes and so is a sensitive probe of the gluon
structure function in the nucleon and its modification in nuclei.
A study of heavy flavor production in p+p and d+Au collisions in
various kinematic regions presents an opportunity to probe cold
nuclear medium effects; parton shadowing, color glass condensate,
initial state energy loss, and coherent multiple scattering in
final state interactions. The PHENIX muon arms cover both forward
and backward directions in the rapidity range of $1.2 < |\eta| <
2.4$. We investigate single muon production from open heavy flavor
and light mesons decay in p+p and d+Au collisions at forward and
backward rapidity.
\end{abstract}


\section{Introduction: }

Heavy quarks are believed to be mostly created from initial gluon
fusion in hadronic collisions. Since they are massive, heavy
flavor hadrons are proposed to be ideal probes to study the early
stage dynamics in heavy-ion collisions.

Measurements of heavy quark production in p+p interactions at
collider energies serve as important tests for perturbative
Quantum Chromo Dynamics (pQCD), while measurements in d+Au
collisions serve to calibrate the effects of the cold nuclear
medium. Both observations create a very important baseline for
understanding the hot-dense matter created in Au+Au collisions.
Since the initial formation of open and closed charm are both
sensitive to initial gluon densities, open charm production serves
as an appropriate normalization for $J/\psi$ production.

In $\sqrt{s_{NN}} = 200$ GeV d+Au collisions at RHIC, measurements
at forward rapidity (deuteron direction) probe the shadowing
region with momentum fractions in Au near $x_2 = 0.01$ while the
anti-shadowing region is probed in backward rapidity (Gold going
direction) with momentum fractions in Au near $x_1 = 0.1$. Recent
models of gluon shadowing~\cite{glouonshadowing}, color glass
condensate~\cite{CGC} and recombination~\cite{recombination} are
implemented to understand the hadron and open charm production at
forward rapidity. All of these three models predict suppression in
the small $x$ region. It is very important to have precise
measurements of open heavy flavor and hadron production to
disentangle these different models.

\section{Experimental technique}

The PHENIX experiment~\cite{phenix} has measured open charm
production through observation of semi-leptonic decays at forward
and backward rapidity with the PHENIX muon spectrometer. The
PHENIX muon arms cover both forward and backward directions in the
rapidity range of $1.2 < |\eta| < 2.4$, which covers both
shadowing and anti-shadowing regions.

The decay of heavy flavor is prompt, and produces a track with an
origin at the collision vertex.  Another source of prompt
muon-like tracks is hadrons that punch through the shielding in
front of the muon spectrometer.  The acceptance of the
spectrometer for these prompt tracks is relatively independent of
the vertex location $z$.  Another background source is the weak
decay of those same hadrons before reaching the shielding -- these
non-prompt muon tracks have an origin separated from the collision
vertex, and the acceptance for these tracks is strongly
$z$-dependent.

A PYTHIA simulation shows around 75$\%$ of prompt muons with $p_T
>$ 0.9 GeV/$c$ come from open charm decay in d+Au collisions, while 11$\%$
come from open bottom decay. Prompt muons are produced close to
the collision vertex. We can separate heavy flavor decays and
light hadron decays experimentally by studying the shape of the
vertex distribution.

The normalized event vertex distribution of reconstructed muons is
given by

\begin{equation}
\frac{1}{N_{measured}^{MB}}\frac{d^3N(z,\eta,P_T^{\mu^{\pm}})}{dzdP_T^{\mu^{\pm}}d\eta}
\propto \{ \alpha(P_T, \eta) (z-z_{eff}^0) + \beta(P_T, \eta) \}
\end{equation}
where $z$ is the event vertex and $z_{eff}^0=\pm 41$cm for north
and south arms, respectively. $\alpha(P_T,\eta)$ and $\beta(P_T,
\eta)$ are determined from event vertex distribution. The
distribution of detected muons from light hadron decay is
collision vertex dependent due to the acceptance of the muon
spectrometers, and this is described by the $\alpha$ parameter.
The prompt muon and the hadron punch through tracks will
contribute to the vertex independent part and this is described by
the $\beta$. The data were used to subtract the hadron punch
through. The PHENIX muon spectrometer includes a "muon identifier"
consisting of thick iron sheets with detectors in the gaps between
the sheets.  Muons are more likely to penetrate into the deeper
gaps than are hadrons. By analyzing the hadrons stopped at gap 2
and 3, one can estimate the hadron contribution at gap4 using an
attenuation model~\cite{run2pp}. With this statistical method, we
can measure the yield of muon from heavy flavor decays with the
PHENIX muon arms.

\begin{figure}[ht]
\begin{center}
\epsfig{figure=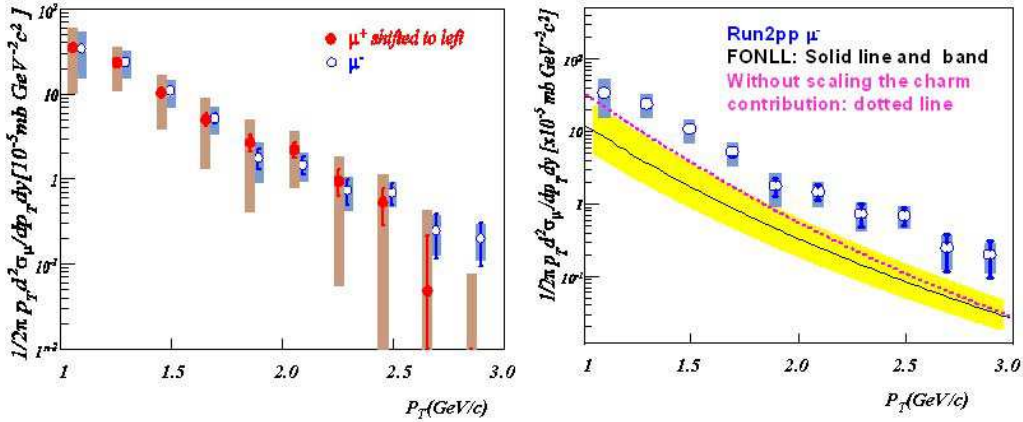,height=3.5in} 
\vskip -3.3cm \caption{Left: $p_T$ spectrum of prompt muons. Error
bars indicate statistical errors and shaded bands indicate
systematic errors. Right: $p_T$ spectrum of negative prompt muons
from Figure 1 with PYTHIA calculations. \label{ppcharm} }
\end{center}
\end{figure}\vskip -1cm

\section{Open Charm results for p+p collisions}

The invariant differential cross section for prompt muon
production at forward rapidity $(1.5<\eta<1.8)$ has been measured
by the PHENIX experiment over the transverse momentum range $1
<p_T < 3 $ GeV/$c$ in $\sqrt{s_{NN}}$ = 200 GeV p+p collisions at
the Relativistic Heavy Ion Collider.

The resulting muon spectrum from heavy flavor decays is compared
to PYTHIA and a next-to-leading order perturbative QCD calculation
showing in Figure 1. PHENIX muon arm data (at forward and backward
rapidity) is compatible with the PHENIX charm measurement at y = 0
~\cite{singleepp}, and it exceeds predictions from PYTHIA and
FONLL.

\section{Nuclear Modification Factor in d+Au collisions}

The nuclear modification factor of d+Au collisions is defined as
the particle yield per nucleon-nucleon collision relative to the
yield in p+p collisions. The nuclear modification factors with
muons from light hadron decay and from heavy flavor decay are
shown in Figure 2 and Figure 3.
\begin{figure}[ht]
\begin{center}
\epsfig{figure=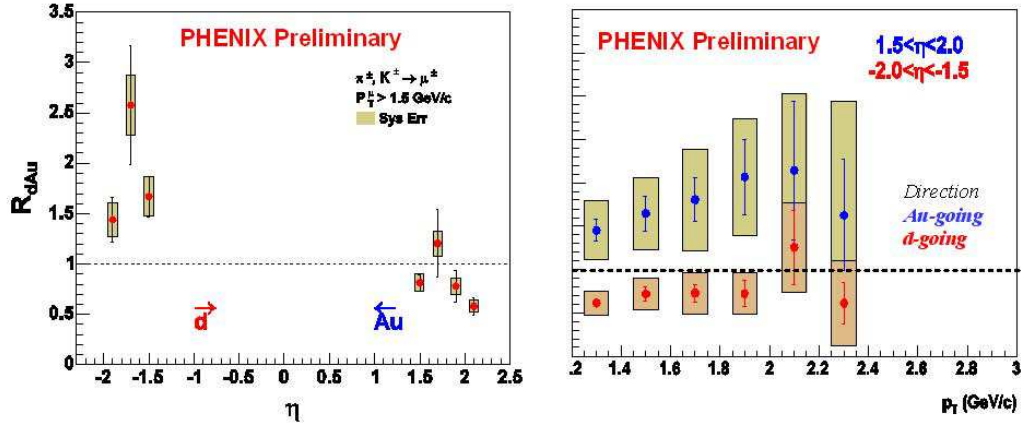,height=3.0in} 
\vskip -1.9cm \caption{Left: $R_{dAu}$ for hadrons vs $\eta$;
Right: $R_{dAu}$ for hadrons vs $p_T$. South Arm is Au-going
direction and North Arm is deuteron-going direction.
\label{rdau_hadron} }
\end{center}
\end{figure}
\begin{figure}[ht]
\begin{center}
\epsfig{figure=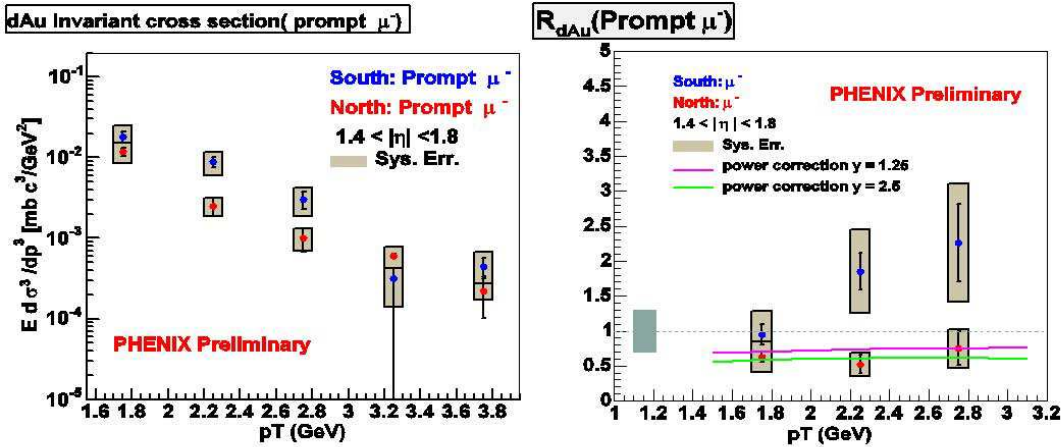,height=2.8in} 
\vskip -1.6cm \caption{Invariant spectra of the prompt muons
(left) and nuclear modification factor of prompt muons (right) in
d+Au collisions. The theoretical curves are come from power
correction model at $\eta = 1.25$ and 2.5
\label{rdau_charm}~\cite{ivan}. }
\end{center}
\end{figure}
Muons from both light hadron decay and prompt single muon
production show suppression at forward rapidity and enhancement in
the backward direction.

\section{Summary and Outlook}

FONLL and PYTHIA 6.205 under-predict the prompt $\mu$ yield at
forward rapidity in p+p collisions at 200 GeV/$c$. We observe a
significant cold nuclear medium effect in forward and backward
rapidity in d+Au collisions at 200 GeV/$c$. For both muons from
open heavy flavor decays and light hadron decays, a suppression in
forward rapidity is observed. It is consistent with CGC and power
correction model~\cite{ivan}. The mechanism of the observed
enhancement at backward rapidity needs more theoretical
investigation. Anti-shadowing and recombination could lead to such
enhancement.

We need a more precise d+Au measurement to understand the cold
nuclear medium effects as a baseline for understanding the hot
dense matter produced in Au+Au collisions.

\end{document}